\documentclass[prd,showpacs,amsmath,showkeys,twocolumn,floatfix]{revtex4} 
\usepackage{epsfig,dcolumn}
\usepackage{graphicx}
\usepackage[usenames]{color}
\usepackage{graphicx}
\usepackage{bm}

\def\lsim{\mathrel{\rlap{\lower4pt\hbox{\hskip1pt$\sim$}}
    \raise1pt\hbox{$<$}}}
\def\gsim{\mathrel{\rlap{\lower4pt\hbox{\hskip1pt$\sim$}}
    \raise1pt\hbox{$>$}}}
\begin{document}



\title{ $P$ wave $\pi\pi$ amplitude from dispersion relations} 

\author{  Adam P. Szczepaniak, Peng Guo}
\affiliation{ Physics Department and Center for Exploration of Energy and Matter \\ 
Indiana University, Bloomington, IN 47405, USA. }

\author{  M. Battaglieri, R. De Vita} 
\affiliation{Instituto Nazionale di Fisica Nucleare, \\ 
 Sezione di Genova, 16146 Genova, Italy }

\date{\today}

\begin{abstract} 

We solve the dispersion relation  for the $P$-wave $\pi\pi$ amplitude.We discuss the role of the 
 left hand cut vs  Castillejo-Dalitz-Dyson (CDD),
 pole contribution and compare the solution with a generic quark model description. We review the the generic properties of   analytical partial wave scattering and production amplitudes and discuses their applicability and  fits of experimental data. 

  \end{abstract} 

\pacs{11.80.Et,11.55.Fv,13.25.-k,14.40.Be}

\maketitle

\section{Introduction} 
\label{sec-0} 
 There is a revival of interest in  the light hadron spectroscopy. 
  Several experiments  are currently carrying out searches for new and  exotic resonances and more facilities are planed for the near future.  
 This is because the current knowledge of the spectrum is still very poor; in particular in the meson sector the Particle Data Group~\cite{PDG} lists approximately a 
 dozen or so of  well established resonances and a plethora of states with poorly determined characteristics whose existence, in some cases, is even questionable.
 Establishing resonance  parameters requires partial wave analysis and in particular knowledge of  amplitudes in an unphysical region of the energy. 
 This necessitated use of methods, such as  dispersion relations, that explore analytical properties of the physical amplitudes.  Since the unitarity condition
  relates the measured cross sections to the 
   discontinuities of  the amplitude, dispersion relations can be solved to obtain amplitudes for any complex (unphysical) value
     of the energy.  Unfortunately, even in the simplest cases, no complete information about discontinuities  of the amplitude is available and 
    approximations must be devised to solve the dispersion relations.  Nevertheless, application of dispersion relations  can lead to precision studies of resonance parameters, as was recently shown to be the case  of the $\sigma$ resonance in $S$-wave $\pi\pi$ scattering~\cite{Oller:1997ng,Oller:1998zr,Pelaez:2003eh,Caprini:2005zr,Kaminski:2006qe,Caprini:2008fc}. 
    Not only that but, exploring the relation between chiral 
     effective field theory and   QCD parameters, it was possible to shed new light onto 
      the resonance structure~\cite{Pelaez:2003dy,Pelaez:2006nj,GarciaRecio:2003ks,Guo}. 
 Further insight into the microscopic nature of a resonance  can be obtained 
  by studding its shape in different production process~\cite{Aitchison:1970tp,Aitchison:1969tq,Aitchison:1977sm}.  For example, a spatially extended  source  that  produces  a pair of mesons  is not expected to couple  strongly to a resonance that  is  dominated by valence quarks, {\it i.e.} whose  wave function is spatially compact. 
  Thus a compact source such as charmonium, when  decaying into light mesons,  is expected to couple to short-range components of intermediate states and resonances.

      It  thus follows that  in partial wave analysis it is important to use amplitudes 
   with  proper analytical behavior across both the physical (right) and unphysical (left) cut. The discontinuity of the amplitude across  the right hand cut ({\it rhc}) is constrained by unitarity and  relates partial wave amplitudes in production and  scattering reactions. The latter  "contains" resonances and these 
    will also  appear in the production process albeit with modified characteristics, that depend on the nature of production dynamics. The  left hand cut ({\it lhc}) determines the production 
     dynamics  and may enhance or suppress 
      a particular resonance. 
   For example,   
     the QCD nature  of the $a_1$  as a member of the quark multiplet 
      was established by application of analytical methods to the $3\pi$ production  in $\pi p$ collisions~\cite{Bowler:1975my}. More recently,   an analysis of the $\pi\pi$ production in $\gamma\gamma$ 
       fusion based on dispersion relations showed that the $\sigma$ meson is expected to contain a substantial short range component, besides the dominant $\pi\pi$ 
        component~\cite{Pennington:2006dg}.
  The long range component originates from long range  interactions between 
  pions and, only  in the limit of asymptotically large number of colors,  it is excepted to be suppressed compared to the quark component.      
    
            Unfortunately it is a common practice to ignore  the above mentioned intricacies of amplitude analysis and rely on simple  parameterizations 
             {\it i.e.} in terms of a superposition of   Breit-Wigner (BW)  resonances. More sophisticated analyses would use pole parameterizations for the  $K$-matrix 
          and implement unitarity restrictions across the right hand cut, but  the role of the left hand cut is often overlooked. The main motivation of this paper is illustrate its role using as an 
           example 
                 the  $\pi\pi$ $P$-wave  amplitude which is dominated by the well-known $\rho(770)$ resonance.  We  will show how the  general analytical properties of amplitudes discussed above emerge in this particular case 
               and how the behavior of the amplitude on the {\it lhc} can 
               modify the resonance production.
  
  The paper is organized as follows. In the next section we summarize the analytical properties of  partial wave amplitudes, both in the case of scattering and production. In  Section~\ref{sec-2} we solve the dispersion relation for the $P$-wave amplitude and discuss a simple parameterizations 
   showing how  the shape of the $\rho$-meson changes depending on the characteristics of the production process. 
    Summary and outlook are given in Section~\ref{sec-3}. 
 
 \section{ Analytical  properties of  scattering and production partial wave amplitudes} 
 \label{sec-1} 
 In this work we focus on partial wave  amplitudes which are a function of a single energy variable. 
This applies, for example to  photoproduction of  pion pairs on the nucleon, $\gamma p \to \pi^+ \pi^- p$.  At fixed photon energy $E_\gamma$,  momentum transfer, $t$  between the target and recoil nucleon and photon and nucleon helicities $\lambda_i$,  the di-pion production amplitude, $A$  is a function of the di-pion invariant mass squared $s$,  and the spherical angle $\Omega =(\theta,\phi)$ which 
  describes the direction of motion of   the $\pi^+$ in the di-pion rest frame, 
  \begin{equation} 
  A  = A(s,\Omega, E_\gamma,t,\lambda_i). 
  \end{equation} 
  The angular dependence can be expanded in a series of partial waves labeled by the spin 
    of the di-pion pair, $l$ and its projection on,  {\it e.g.} the photon direction ( $t$-channel helicity frame) 
    \begin{equation} 
    A = \sum_{lm} A_{lm}(s;E_\gamma,t) Y_{lm}(\Omega).
    \end{equation} 
    Finally the charged di-pion state can be decomposed into states of total isospin, $I=0,1,2$ 
    \begin{equation} 
    A_{lm} = \sum_{I,I_3} A_{lm,II_3} \langle 11,1-1|II_3\rangle.
    \end{equation} 
The partial wave analysis can now be performed in bins of $E_\gamma$ and $t$  for  a set of  partial wave amplitudes $A_{lm,II3}$  of a singe variable $s$ -- the di-pion mass squared. The singularities of these amplitudes in the complex-$s$ plane will in general 
  depend on $t$.  This parametrization was recently used in the analyses of the two pion 
  photoproduction data from CLAS at JLab~\cite{Battaglieri:2008ps,Battaglieri:2009fq}  and  is currently underway for the hadronic decays of light charmed hadrons using CLEO and BES 
  data~\cite{CLEO}.

   In the following we will concentrate on production of  pion pairs. The case of heavier 
     mesons and/or  baryons  can be formulated analogously, albeit with complications arising from presence of sub-threshold cuts and spin. 
       As a function of the  di-pion invariant mass  squared, $s = 4 (q^2 + m^2)$, with $q$ being the relative momentum, and at fixed values of other kinematical variables, the $l$-th  partial wave production amplitude is  factorized into the angular momentum barrier factor,   $q^{l}$ and the amplitude $F(s)$ that is 
      free from kinematical singularities
      \begin{equation}
      A(s) = q^l F(s). \label{A}  
      \end{equation}
       The amplitude $F(s)$  is a real analytical function with a right hand cut  
   starting  at threshold, $s_{th} = 4m^2$ and with each open inelastic channel, $s_i> s_{th}$ 
  contributing to the discontinuity across the cut.  
 The value of this discontinuity    is  constrained by unitarity. For  $s$ on the positive real axis, defining $F^\pm(s) \equiv F(s \pm i \epsilon)$  as the boundary value of the  function $F(s)$ on the upper (lower)  lip of the cut, unitarity implies
  \begin{equation} 
 Im F^+(s) = t^-(s) \rho(s)  F^+(s) \theta(s - s_{th})   + \sigma(s) \theta(s - s_i). \label{Frhc} 
\end{equation} 
Here $t^-(s) = t(s - i\epsilon)$ is the boundary value on the lower lip on the right hand cut of the 
$\pi\pi$ l-wave scattering amplitude $t(s)$ considered as an analytical function of $s$. The two body phase space is given by  $\rho = (1- s_{th}/s)^{1/2}$ and $\sigma(s)$ represents the contribution from production of inelastic  channels above $s_i$.  We are using the following normalization of the $\pi\pi$ amplitude, 
 \begin{equation} 
\eta(s) e^{2 i\delta(s)} = 1 + 2 i \rho(s) t^+(s),  
 \end{equation} 
 where $\delta, \eta$ is the phase shift and inelasticity, respectively. 
 In addition to the right hand unitarity cut, the production amplitude is discontinuous  for negative $s$. 
 The location of the left hand cut  $s < s_0 \le 0$ depends on the underlying  production 
  dynamics. It is related to  thresholds for particle production in the  crossed channels and plays the role of the driving term in the integral equation for the amplitude that follows from a dispersion relation.  It has a similar role to that of the potential in the non-relativistic 
  Lippmann-Schwinger equation of the  Schr\"odinger theory. 
   Thus for $s$ real and $s  < s_0$, 
 \begin{equation} 
 Im F^+(s) = Im F_L(s + i\epsilon) \theta(s_0 - s),  \label{Flhc} 
 \end{equation} 
 with  the  "potential" $F_L(s)$ defined as a real analytic function in the complex $s$-plane 
 with the {\it lhc}  discontinuity given by Eq.(\ref{Flhc}), 
 \begin{equation} 
 F_L(s) = \frac{1}{\pi} \int_{-\infty}^{s_0} ds_L \frac{ Im F^+(s_L) }{s_L - s}.  
 \end{equation} 
 From Eqs.(\ref{Frhc}),(\ref{Flhc}) it follows that $F(s)$ satisfies an integral equation, 
 \begin{eqnarray} 
 F(s) = F_L(s) & + &  \frac{1}{\pi} \int^\infty_{s_{th}} ds_R \frac{ t^-(s_R) F^+(s_R)}{s_R-s}  \nonumber \\
  &+ & \frac{1}{\pi} \int_{s_i}^\infty ds_R\frac{ \sigma(s_R) }{s_R-s }, \label{Fdisp}
 \end{eqnarray}
  where we assumed that the integrands vanish as $|s| \to \infty$. If  $F(s)$ is bound by a polynomial in $s$, the dispersive integral can be  made convergent by subtractions. These introduce additional parameters, that are related to the asymptotic behavior of the amplitude. In general  little is known about the left hand cut discontinuity, {\it i.e} $F_L(s)$.  For example in bootstrap  calculations it is approximated by particle, or more generally Regge exchanges with parameters adjusted so that the solution of the dispersion relation reproduces the  known resonances. More recently chiral effective field theory has been used to construct approximation to $F_L$ at low energies  by expanding it in powers of $s/\Lambda_{\chi}^2$, where  $\Lambda_{\chi} = 4\pi f_\pi$ is the chiral scale~\cite{Pelaez:2003eh,Caprini:2005zr,Truong:1988zp-2,Dobado:1996ps,Dobado:1996ps-2}.
    The dispersion relation in  Eq.(\ref{Fdisp}) is an integral equation for the production amplitude $F(s)$ which as input takes {\it i})   the scattering amplitude $t$, {\it ii}) the amplitude representing contributions from production of inelastic channels $\sigma$ and {\it iii}) the "potential", $F_L$. 
   The analytical solution of Eq.(\ref{Fdisp}) is known,  and involves the function $D(s)$ from the 
  $N/D$ decomposition of the scattering amplitude, $t(s)$~\cite{CM,Bjorken:1960zz,FW}.  Even though the $N/D$  method for solving partial wave dispersion relations  has been extensively studied in the past, due to its  important role in partial wave analysis  we summarize its main features in the following paragraphs.

   \subsection{ N/D representation of the scattering amplitude} 
    The dispersion relation for the partial wave scattering amplitude $t(s)$  is given by 
   \begin{equation} 
   t(s) = \frac{1}{\pi} \int^{s_0}_{-\infty} ds_L \frac{Im t^+(s_L)}{s_L - s} + \frac{1}{\pi} \int_{s_{th}}^\infty  ds_R \frac{ Im t^+(s_R)}{s_R - s}.
    \label{t1}
   \end{equation}
It becomes an integral equation for $t(s)$, once unitarity is implemented. It relates  the {\it rhc} discontinuity in $t$ to the amplitude itself via, 
    \begin{eqnarray} 
   Im t^+(s_R) &= &  \rho(s_R) |t(s_R)|^2 \theta(s_R - s_{th})  \nonumber \\
    & + &  \frac{1 - \eta^2(s_R)}{4}\theta(s_R - s_i).  \label{trhs} 
   \end{eqnarray} 
   The solution of Eqs.(\ref{t1}),(\ref{trhs}) is expressed in the form, 
   \begin{equation} 
   t(s) = (s - s_{th})^l \frac{N(s)}{D(s)} \label{NDorg}, 
   \end{equation}
   where the angular momentum barrier factor has been explicitly factored out and $N(s)$, 
    and  $D(s)$ are defined so to have only left and  right hand  
    cut, respectively.  The  definition of $N$  and $D$ is unique up to an overall constant, which we fix by normalizing $D$ at the elastic threshold, 
      $D(s_{th}) = 1$. For $l\ge 1$ the uniqueness follows from the threshold   behavior, $t(s_{th}) 
       =  0$ and the asymptotic behavior at large $s$, 
       $|t( \infty + i\epsilon)| < O(1)$. For $S$-waves, one subtraction  may be needed to make the integrals in Eq.(\ref{t1}) convergent and 
        consequently  the dispersion relation for $t$ (or $N$ or $D$) contains one undetermined constant ({\it i.e.} scattering length). 
        
          From now on we focus on the $P$-wave, $l=1$ amplitude. From the dispersion relation for $t$ it  follows that,  
    \begin{eqnarray} 
    & & N(s) = \frac{1}{\pi} \int_{-\infty}^{s_0} ds_L \frac{ Im t(s_L) D(s_L)} {(s_L - s_{th}) (s_L - s) }  \nonumber \\
    & & D(s) = 1 - \frac{s - s_{th} }{\pi} \int^{\infty}_{s_{th}} ds_R  \frac{\rho(s_R) N(s_R) R(s_R) }{s_R - s} ,  \nonumber \\  \label{ND} 
    \end{eqnarray} 
where $R \equiv  Im t^+/\rho |t|^2$,  differs from unity for $s_R > s_i$  because  of inelastic channels 
 opening.   
 Substitution of the equation for $N$ into the equation for $D$ leads to an integral equation, 
\begin{equation} 
D(s) = 1 - \int^{s_0}_{-\infty} ds_L  K(s,s_L) D(s_L)  
\end{equation} 
where the kernel 
\begin{equation} 
K(s,s_L) = \frac{ s - s_{th} }{\pi^2}  \int_{s_{th}}^\infty  ds_R  \frac{ \rho(s_R) R(s_R)  Im t(s_L) }{(s_L - s_{th})(s_L - s_R)(s_R - s)}, \label{K} 
\end{equation} 
is given in terms of the "potential",  {\it i.e}  discontinuity of $t$ across the left hand cut, and the inelastic contribution proportional to  $R-1$. 
 Since  $|t( \infty + i\epsilon)| < O(1)$,  
 it follows from Eq.(\ref{ND})  that asymptotically  $|N(s)|  < O(1/s)$  and $|D(s)| < O(1)$ and no subtractions are needed. 

The dispersion relation for $D$ in Eq.(\ref{ND}) assumes $t \ne 0$, which is always the case  in the inelastic region when $\eta \ne 1$. 
 Otherwise zeros of $t$ correspond to poles of $D$, the so called CDD poles that are not accounted for by Eq.(\ref{ND})~\cite{CDD}. Therefore in presence of zeros of the scattering amplitude,  the CDD poles have to be added "by hand"  and result in, 
   \begin{eqnarray}
 & &   D(s) = 1 -  \frac{s- s_{th}}{\pi}  \int_{s_{th}}^\infty  ds_R \frac{\rho(s_R) N(s_R) R(s_R) }{  s_R -s }  \nonumber \\
&&  -  (s - s_{th})  \Pi_{p=1}^{N_p} \frac{\gamma_p}{ (s_p - s ) (s_p - s_{th})}, \nonumber \\ \label{DCDD} 
 \end{eqnarray} 
 where $\gamma_p$ and $s_p$ are the (real)  residue and position of the $p$-th CDD pole, respectively.  At every CDD pole the phase of the amplitude passes through $180^0$    and 
  Levinson theorem relates the number of CDD poles to the phase shift at infinity, 
         $\delta(\infty)/\pi   = N_p $~\cite{Lev} .   If the residue of a CDD pole  is small, then $D(s)$ will develop a zero on the second sheet near the position of the pole, {\it i.e.}
         will  produce a resonance. Thus, in the past  it has been proposed to  identify 
          CDD poles with the elementary quark bound states that turn into  physical resonances when coupled to the continuum channels.  
           Indeed it has been shown that in potential models describing, for example  scattering off  a static source with internal structure,  CDD poles in the 
            dispersion relation for a scattering amplitude correspond to excitations of the target. 
  There is, however no proof  of such correspondence in QCD,  and as we will show in the following section, a  correspondence  between CDD poles and quark model states may be more complicated than in potential theory. 
          
It follows that, even though unitarity and the left hand cut discontinuity do not yield a unique solution of the dispersion relation for  the  scattering amplitude $t$,  the CDD ambiguity disappears  if the phase shift and inelasticity are known.   
 In this case, the solution of Eq.(\ref{DCDD})  is given by, 
   \begin{equation} 
   D(s) = \Pi_{p=1}^{N_p}\left[  \frac{s_{th}  - s_p  }{s - s_p} \right] \Omega_\phi(s),  \label{OO} 
   \end{equation} 
   where the first factor comes from the CDD poles and the Omnes-Muskhelishvili  function $\Omega$~\cite{M,O}
\begin{equation}    
\Omega_\phi =     \exp\left( - \frac{s - s_{th}}{\pi} \int^\infty_{s_{th}} ds_R \frac{ \phi(s_R) } {(s_R - s)(s_R - s_{th}) } \right) 
   \end{equation} 
   is given in  terms  of the phase of $t$, $t = |t| \exp(i \phi)$, which is identical to the phase shift $\delta$ in the elastic region. Outside the elastic region the phase $\phi$ is determined from phase shift and inelasticity and is bound between $0$ and $\pi$. 
   
   \subsection{ N/D representation of the production amplitude} 
   
 The solution of the dispersion relation, Eq.(\ref{Fdisp}) for the production amplitude can be 
  represented in a number of equivalent ways~\cite{Pham:1976yi}.    From the point of view of data parametrization, a particularly useful representation is given by 
%
\begin{equation} 
F(s) = \frac{G(s)}{D(s)} = \frac{ F(s_{th})  + G_L(s) - G_i(s)} {D(s)},  \label{F1} 
\end{equation} 
where $D(s)$ is given by Eq.(\ref{OO}) (or (\ref{DCDD})), the numerator function $G_L$ contains only the left hand cut, 
\begin{equation} 
G_L(s) =   \frac{s - s_{th}}{\pi} \int_{-\infty}^{s_0} ds_L \frac{ Im F_L(s_L) D(s_L) }{(s_L - s)(s_L - s_{th})} 
\end{equation} 
while the function $G_i(s)$ describes inelastic contribution to production, 
\begin{equation}
G_i(s) =    \frac{ s-s_{th}}{\pi } \int_{s_i}^\infty ds_R \frac{  Im \sigma(s_R) D(s_R) }{
t^-(s_R)  \rho(s_R)(s_R - s)(s_R - s_{th})  }. 
\end{equation} 
We have assumed that at most one subtraction is needed, which is the case if $D$ has a CDD pole at infinity that as we will discuss later is the likely case for the $P$-wave. 
 Expression for the production amplitude in Eq.(\ref{F1})  can easily be adopted to fits of experimental data. Assuming that $D(s)$ is a known function (we will discuss special cases in the  next  Section),  the numerator in Eq.(\ref{F1}) depends on the inelastic contribution ($\sigma$) and  the production dynamics ($F_L$). These are the quantities one should  extract from  the data. The form in Eq.(\ref{F1}) suggests that, instead of parametrizing   
   $F_L$ and $\sigma$,  it would be more efficient to parametrize the entire expression appearing in the  numerator. It represents a sum of two analytical functions one with the left an the other with a right hand cut that begins at the first inelastic threshold. Thus  one can conveniently use a 
   series expansion, 
      \begin{equation} 
   G_L(s) = \sum_{i=0} c_{i,L} z_L^i(s), \; \; G_i(s) = \sum_{i=0} c_{i,R} z_R^i(s), \label{conf1} 
   \end{equation} 
   where $s \to z_{L(R)}(s)$  represents a conformal mapping of a cut plane on a unit 
   circle~\cite{Yndurain:2002ud}, 
   \begin{eqnarray}
& &  z_L(s) = \frac{  1  - \sqrt{1 - s/s_L}   }{1 + \sqrt{1 - s/s_L} }  \nonumber \\
& & z_R(s) = \frac{  1  - \sqrt{1 - s/s_i}   }{1 + \sqrt{1 - s/s_i} }.  \label{conf2}
 \end{eqnarray}
 One would then determine the coefficients $c_{i,L}, c_{i,R}$ from fitting $F(s)$ to the data. 

Another way of representing the solution of  Eq.(\ref{Fdisp}), which in particular amplifies the spacial characteristics of the production process, is obtained by representing  $D(s_L)$ in $G_L(s)$ through its dispersion relation ({\it c.f.} Eq.(\ref{DCDD})), 
 resulting in 
 \begin{eqnarray} 
& &  F(s) =  F_L(s) + \frac{ F(s_{th}) - F_L(s_{th}) }{D(s)}  \nonumber \\
& & -  \frac{s-s_{th}}{\pi D(s)} \int^\infty_{s_{th}} ds_R 
\frac{Im D(s_R)  F_L(s_R)}{ (s_R - s)(s_R-s_{th})}      \nonumber \\ 
  &&  -    \frac{ s - s_{th} }{\pi D(s)} 
   \int_{s_i}^\infty ds_R \frac{ Im \sigma(s_R) D(s_R) }{t^-(s_R)  \rho(s_R)( s_R - s)(s_R - s_{th})}.  \nonumber \\ 
    \label{F2} 
 \end{eqnarray}


 Both representation, Eq.(\ref{F1}) and Eq.(\ref{F2}),   below the inelastic threshold $s < s_i$, satisfy 
  the final state interaction theorem, $\mbox{arg }F(s) = -\mbox{arg }D(s)=  \mbox{arg }t(s) $.
 In Eq.(\ref{F2}), the production amplitude is expressed in terms of the production amplitude $F_L$ evaluated in the physical ($s > s_{th}$) kinematics. Parametrizing $F_L$ via the conformal map ({\it c.f.} Eq.(\ref{conf2})), 
 \begin{equation}
 F_L(s) = \sum_i c_i z_L^i(s),  
 \end{equation} 
and replacing the inelastic contribution with the fit function $G_i$   from Eq.(\ref{conf2}) leads to 
\begin{equation}
  F(s) =    \frac{F(s_{th}) }{D(s)}
  + \sum_i c_i \left[ z^i_L(s) - \frac{z^i(s_{th})}{D(s)}  
  + \frac{I_i(s)}{D(s)} \right]   - \frac{G_i(s)}{D(s)}\label{fit2}
\end{equation}  
with the functions $I_i(s)$ given by 
\begin{equation} 
 I_i(s) = -\frac{s - s_{th}}{\pi} \int^\infty_{s_{th}} ds_R \frac{Im D(s_R) z^i_L(s_R)}{(s_R - s)(s_R - s_{th})} . 
\end{equation} 
The asymptotic behavior is encoded in the rate of convergence of $\sum_i c_i  z^i_L$ at the circle $|z_L| =1$. The applicability of   Eq.(\ref{fit2}), in experimental data fit  is  however  based on the assumption that only a few orders in the expansion  in powers of $z_L$ are needed to describe  a 
particular data set, {\it i.e} one wants the fits to be insensitive to the poorly known asymptotic behavior.  

In the elastic region $s_{th} < s < s_i$ from Eq.(\ref{F2}) one finds 
\begin{equation} 
F(s) = e^{i\delta(s)} \left[ F_L(s) \cos\delta(s) - B(s) \frac{ \sin\delta(s) }{\rho(s)}  \right], \label{Wt} 
\end{equation}
where $\delta$ is the elastic phase shift and $B$ is a real function given by 
\begin{eqnarray}
& & B(s) = \frac{F_L(s_{th}) - F(s_{th}) }{t(s) D(s)} \nonumber \\
& & + 
   \frac{s - s_{th}}{\pi t(s) D(s) } P.V. \int_{s_{th}}^\infty ds_R \frac{Im D(s_R) F_L(s_R)}{(s_R - s)(s_R - s_{th})}  \nonumber \\
& & + \frac{ s - s_{th} }{\pi t(s) D(s) } \int_{s_i}^\infty ds_R \frac{ Im \sigma(s_R) D(s_R) }{t^-(s_R) \rho(s_R) (s_R-s)(s_R - s_{th})}, \nonumber \\ \label{B}
\end{eqnarray}
with $P.V.$ standing for the principal value.  In particular, if  at some $s = s_r$ the phase shift passes through a resonance, {\it i.e}  $90^0$,  
 the first term in Eq.(\ref{Wt}) will produce a zero in the amplitude, while the second term will produce a peak there. 
The zero is due to a destructive interference between the direct production of two-particles and 
 their final state interaction. In the case were production is short-ranged, as discussed earlier, the  start of the left hand cut in $F_L(s)$, $s_0$ is  far away from the physical region, $s_0 << s_{th}$,  
 the principal value integral extends over a large interval in $s_R$, $s_R \lsim |s_0|$ and the contribution from the $B$ term in Eq.(\ref{Wt}) is significant. On the other hand, if the production source is diffuse and inelasticity is small the resonance peak may be significantly distorted by the zero from the $\cos\delta$ term.   Thus analyzing the shape of a resonance can shed light on  its production characteristics and thus its nature. 
   We also note that the first term in the {\it rhs.} of Eq.(\ref{B}) contributes if  the dispersion relation for $G(s)$ requires subtraction. It is needed if $D(s \to \infty) \sim O(s)$ {\it i.e} $D(s)$ has a CDD 
  pole at infinity.

  \section{ $P$-wave $\pi\pi$ scattering and production} 
  \label{sec-2}
\begin{figure}[h]
{\includegraphics[height=7cm]{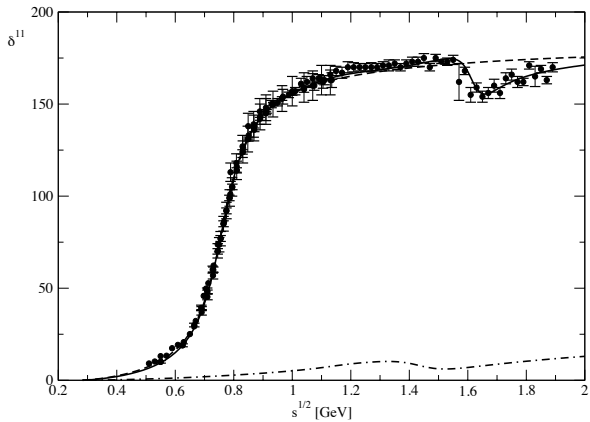}}
{\includegraphics[height=7cm]{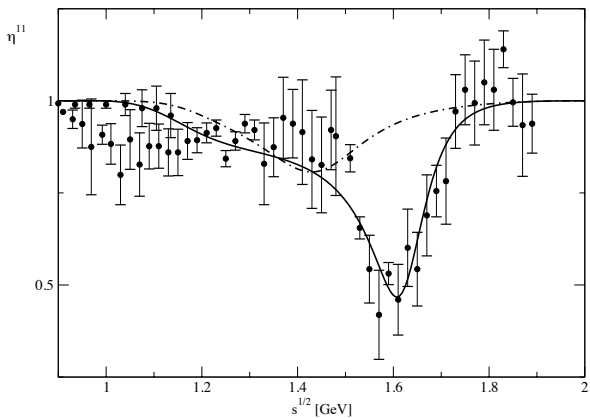}}
\caption{  $P$-wave phase shift (upper panel) and inelasticity (lower panel). Data from ~\cite{Hyams:1973,Protopopescu:1973,Estabrooks:1974}, dahsed-dotted  (solid) line  solution of dispersion relation without (with) a CDD pole. Dashed line is the fit of the quark model from Eq.(\ref{tqm}).} 
 \label{delta} 
\end{figure}

  The solution of the dispersion relation for  the scattering amplitude requires knowledge of the  inelasticity ($R-1$)  and the left hand discontinuity. 
  In the case of the $P$ wave the former is know up to $\sqrt{s} = 1.9 \mbox{ GeV}$ and the latter has been evaluated in Ref.~\cite{Tryon:1974tq}. 
   In the case of $\pi\pi$ scattering, crossing symmetry gives an  additional constraint
    between the "potential"  and the right hand discontinuity. In Ref.~\cite{Tryon:1974tq} 
  a    particular parametrization, for the left hand discontinuity
      \begin{equation} 
      Im t(s) = \frac{a + b s }{s^2} + c [1 - \cos(2\pi x)] 
   \end{equation} 
   with $a = 0.48 \mbox{ GeV}^4$, $b = 1.21 \mbox{ GeV}^2$, $c=0.601$ and $x = (-0.320/(0.283 + s/\mbox{ GeV}^2))^{0.36}$, for  $s \le  -32 m_\pi^2$ 
   has been shown to faithfully represent this constraint.  For $-32 m_\pi^2 < s < s_0 = 0$, 
    crossing  symmetry leads to 
   \begin{eqnarray}
  && Im t(s)  =   \frac{2m_\pi^2}{s - s_{th}} \int_{s_{th}}^{s_{th}-s} ds_R P_1\left( 1 +\frac{2s_R}{s - s_{th}}\right) \nonumber \\
   & & \times \sum_{I=0}^2 C_{1I} \sum_{l=0}^\infty (2l+1) Im t_{lI}(s_R)P_l\left(1 + \frac{2s}{s_R - s_{th}}\right).  \nonumber \\
   \end{eqnarray} 
Since the integration range is limited between threshold and $36 m_\pi^2 \sim 0.675\mbox{ GeV}^2$, the sum can be truncated to include only a few 
 low partial waves and the corresponding amplitudes $Im t_{l,I}$  ($t \equiv t_{1,1}$) expressed in terms of known phase shifts and inelasticities.  For $R$ in Eq.(\ref{ND}) we use the data 
 from \cite{Hyams:1973,Protopopescu:1973,Estabrooks:1974} and assume $\eta = 1$ for $\sqrt{s} > 2\mbox{ GeV}$. The kernel in Eq.(\ref{K}) is  then inverted  numerically 
  and the $D$ function is obtained from computing 
  \begin{equation} 
  D(s) = \int_{-\infty}^{s_0}  ds_L  [1 + K]^{-1}(s,s_L) P(s_L)  \label{inv} 
  \end{equation}  
  where  $P(s_L) = 1$ if no CDD poles are assumed, or $P(s_L) = 1 + \gamma_P (s_L - s_{th})/(s_P - s_{th})(s - s_P)$ for the case of one CDD pole. 
  Since the $P$-wave phase shift stays below $180^0$ all the way up to the first relevant inelastic threshold  associated  with $K\bar K$ production, no
   CDD poles are expected. The exception is a pole at infinity. In our our case, since we have assumed that inelasticity is negligible at large $s$, 
    the CDD pole at infinity could in fact appear at a finite $s_P > (2 \mbox{ GeV})^2$. Once $D(s)$ for $s < 0$ is obtained from Eq.(\ref{inv}), both $N(s)$ and $D(s)$ in the physical region can be computed from Eq.(\ref{ND}) and compared with the measured phase shift and inelasticity. The results are shown in Fig.~\ref{delta}, respectively with the dashed-dotted line showing the result computed without the CDD pole and the solid with one CDD pole, whose residue and pole position have been fitted to the scattering 
    data ~\cite{Hyams:1973,Protopopescu:1973,Estabrooks:1974}. 
It is clear that the resonance behavior corresponding to the $\rho(770)$ is due to the presence of a CDD pole. The exact location of the pole cannot be established due to the unknown behavior of the phase shift at high energies. In the current fit we find the CDD pole with  $\gamma_p = 50.4 \mbox{ GeV}^2$ and $s_p = 6.9\mbox{ GeV}^2$ which for all practical purposes as far as the $\rho$ meson is concerned might be taken to be at infinity.  We also note that in ~\cite{Oller:1998zr} it was erroneously stated that the $P$-wave has a CDD pole at threshold. Vanishing of the amplitude at threshold   is due to the angular momentum  barrier and not 
 to the presence of a CDD pole, the  latter being related 
 to the dynamics of the scattering process.

\begin{figure}[h]
{\includegraphics[height=7cm]{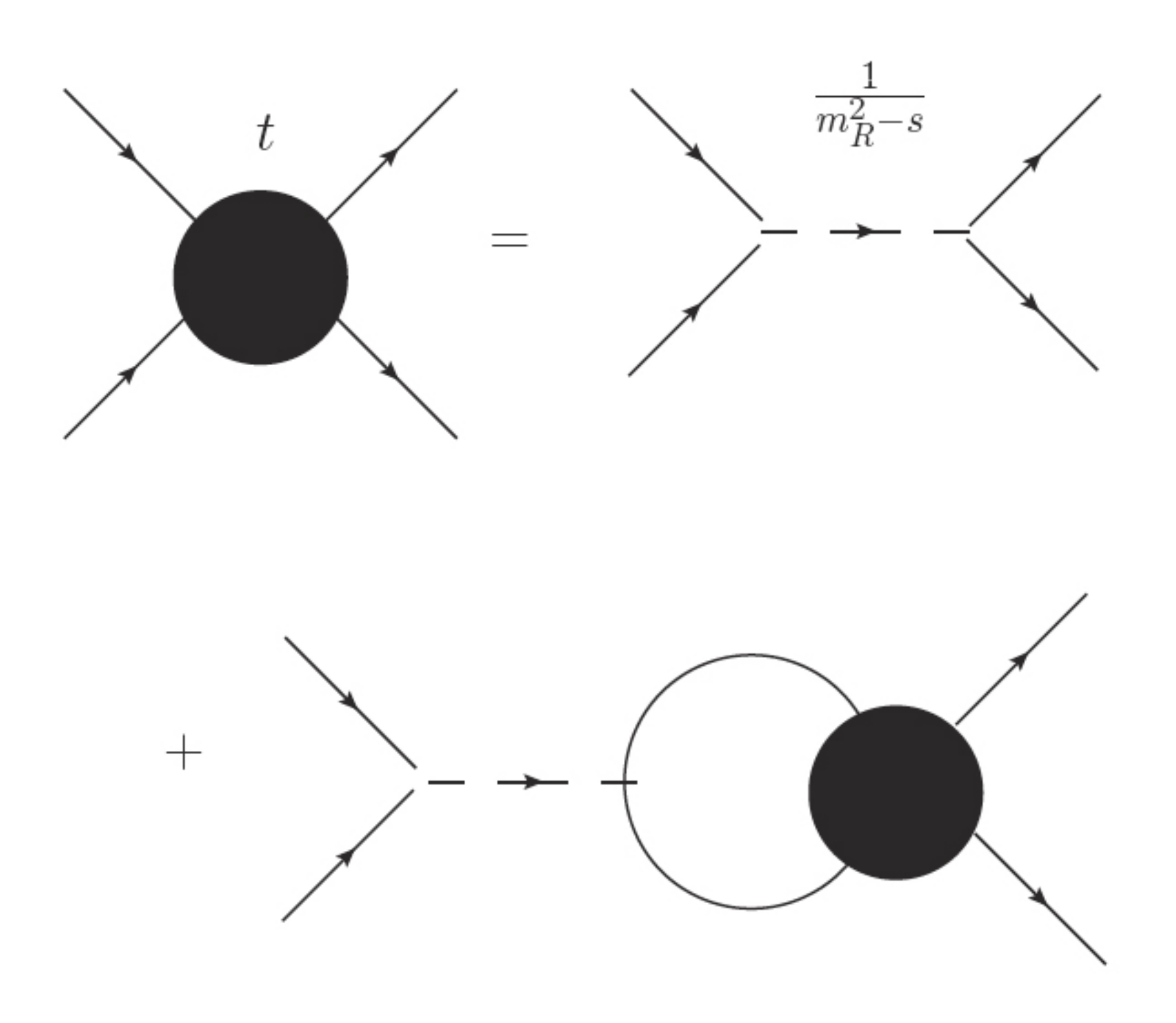}}
\caption{  Quark model representation of the $\rho$ resonance. The bare state with unrenormalised  mass $m_B$  couples to a single, two-particle, 
open channel.} 
 \label{P} 
\end{figure}

\begin{figure}[h]
{\includegraphics[height=7cm]{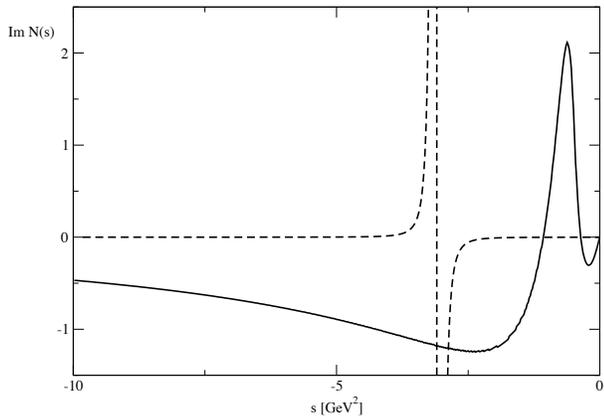}}
{\includegraphics[height=7cm]{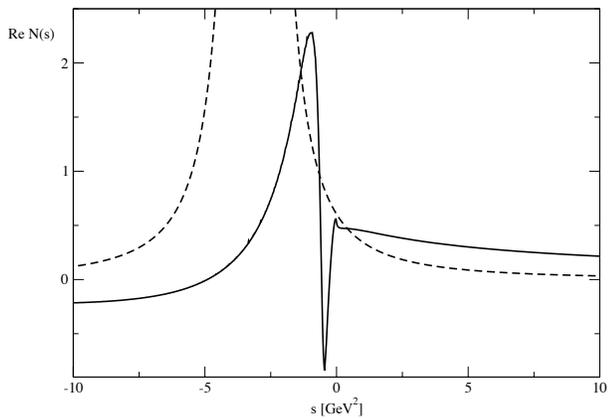}}
\caption{  Imaginary (upper plot) and real (lower plot) of $N$. Solid line is the solution of dispersion relation, Eq.(\ref{ND}) , dashed line of the quark model, 
 Eq.(\ref{tqm}).  } 
 \label{Re-ImN} 
\end{figure}

\begin{figure}[h]
{\includegraphics[height=7cm]{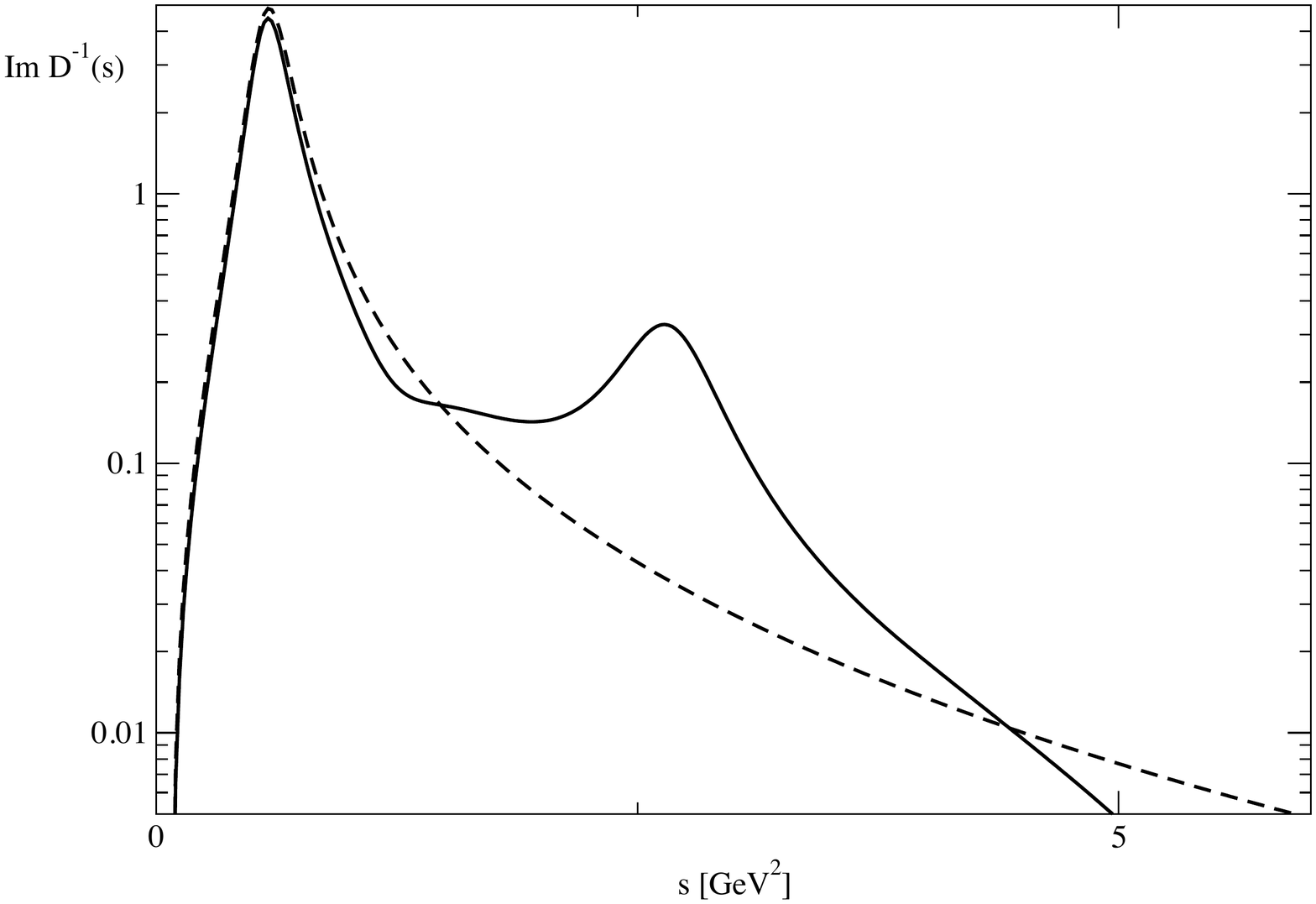}}
{\includegraphics[height=7cm]{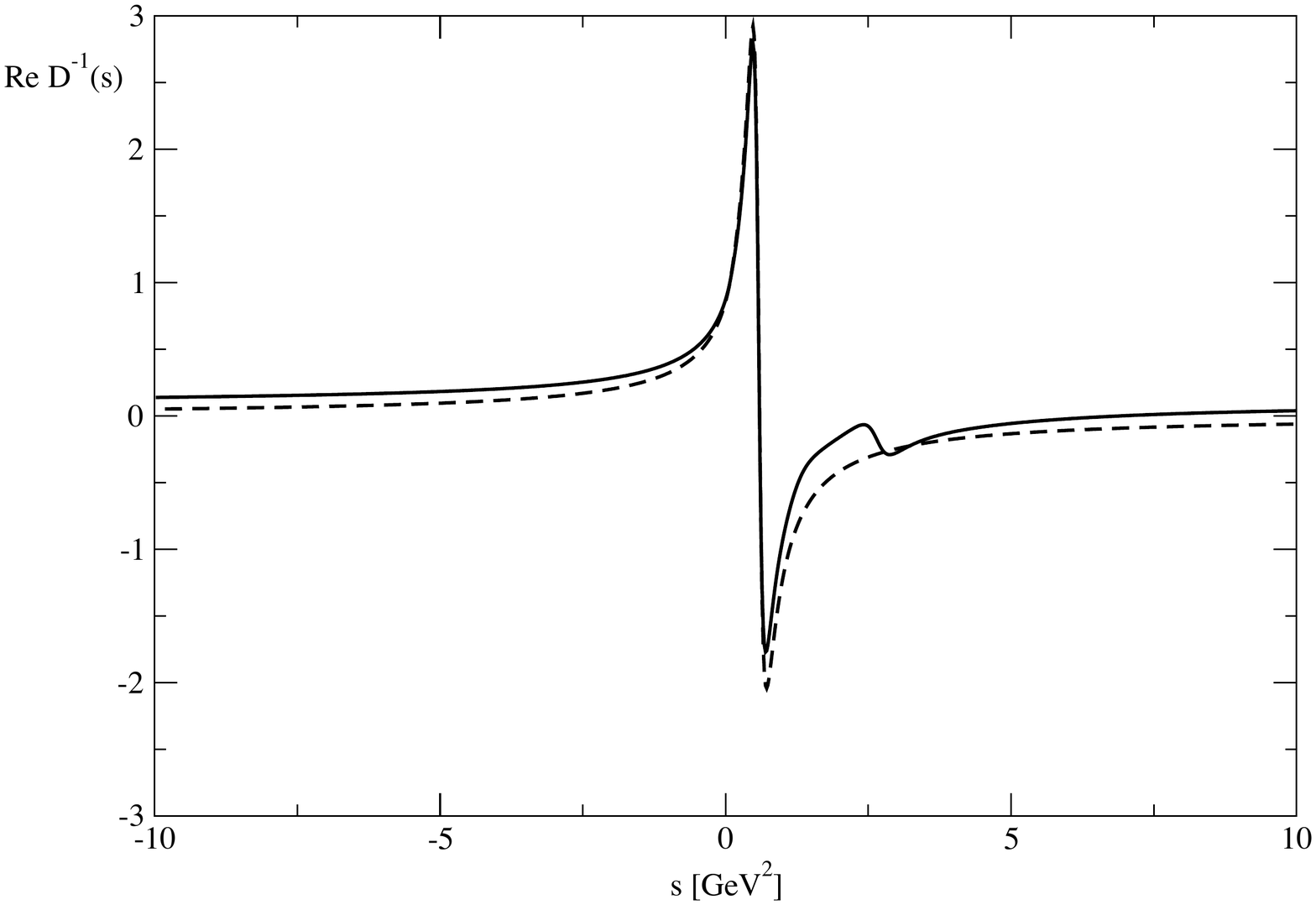}}
\caption{  Imaginary (upper plot) and real (lower plot) of $D^{-1}$. Solid line is the solution of dispersion relation, Eq.(\ref{ND}) , dashed line of the quark model Eq.(\ref{tqm}).  } 
 \label{Re-ImD} 
\end{figure}

\begin{figure}[h]
{\includegraphics[height=7cm]{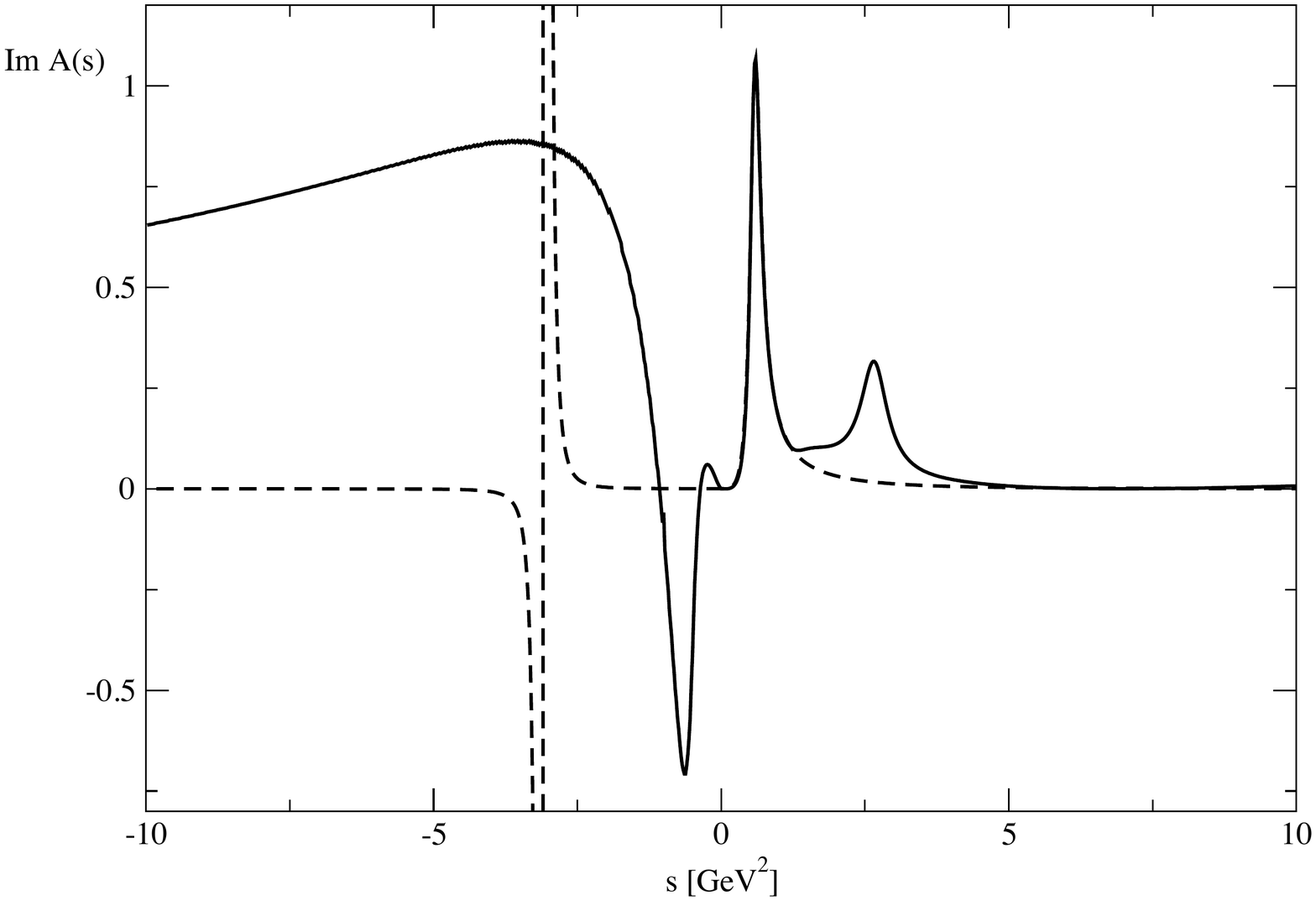}}
{\includegraphics[height=7cm]{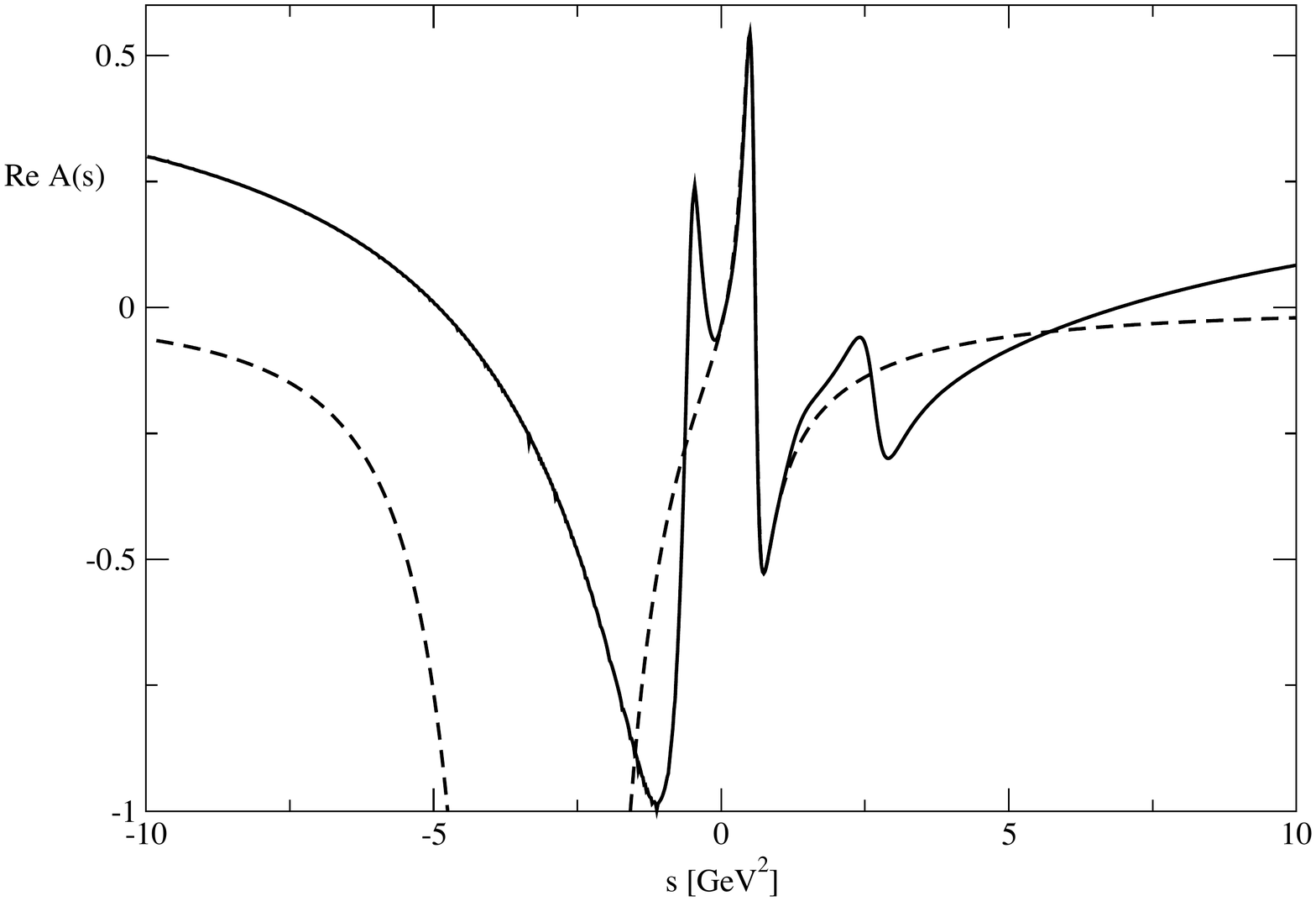}}
\caption{  Imaginary (upper plot) and real (lower plot) of $t$. Solid line is the solution of dispersion relation, Eq.(\ref{ND}) , dashed line of the quark model,  Eq.(\ref{tqm}).  } 
 \label{Re-Imt} 
\end{figure}

 Next explore the connection between the CDD pole description of the $\rho$ resonance with that of the quark model.   From the quark model perspective, the  $\rho$ meson is  considered  as  a quark-antiquark bound state which becomes a resonance when
 coupling to the continuum $\pi\pi$ channel is allowed. Generically, with such dynamical assumptions,  the $P$-wave $\pi\pi$ scattering amplitude can be expressed as a solution of a separable integral equation shown in Fig.~\ref{P} which leads to 
\begin{equation} 
t(s) =  \frac{ (s - s_{th}) f^2(s)}{m_B^2 - s - I_{ff}(s)} \label{tqm} 
\end{equation}
Here $f(s)$ is a vertex  function which represents the coupling of the bare, quark model $\rho$ state to the $\pi\pi$ channel, {\it i.e.} is given in terms of the overlap of the quark model $\rho$ meson  and  the  pion wave  functions.  In general $f(s)$ is analytical in the $s$-plane, except for "potential"-like   singularities  for negative $s$. In the pole approximation it is then given by 
 ($s_f > 0 $), 
\begin{equation} 
f(s) =   \frac{\lambda_f s_f   }{s + s_f}.  \label{fmodel} 
\end{equation} 
The contribution from the open $\pi\pi$ channel results in the modification of the bare $\rho$ propagator $(m_B^2 - s)^{-1}$  by  loop a integral given by 
\begin{equation} 
I_{ff}(s) = \frac{1}{\pi} \int^\infty_{s_{th}} ds_R \frac{(s_R - s_{th}) \rho(s_R) f^2(s_R)}{s_R - s}.  
\end{equation} 
In therms of the $N/D$ representation ({\it c.f} Eq.(\ref{NDorg})) the amplitude of Eq.(\ref{tqm}) is 
becomes, 
 \begin{eqnarray} 
& & D(s)  = 1 - \frac{s- s_{th}}{\pi} \int^\infty_{s_{th}} ds_R \frac{\rho(s_R) N(s_R) }{s_R -  s}  -  (s - s_{th}) C , \nonumber \\ 
& & N(s) = C f^2(s)  \label{NDmodel} 
 \end{eqnarray}
 where the (positive) constant $C$ is defined by 
 \begin{equation} 
 C^{-1} = m_B^2 - s_{th} -  \frac{1}{\pi} \int_{s_{th}}^\infty ds_R  \rho(s_R) f^2(s_R) 
\end{equation} 
Comparing with Eq.(\ref{DCDD}) shows that the quark model representation  of the $\rho$ meson, in the absence of higher excited ($\rho'$) 
 states corresponds to a single CDD pole at infinity {\it i.e.} with  $s_{p},\gamma_p \to \infty$, $\gamma_p/s_p^2 = \mbox{ fixed}$.  We fit the three constants $s_f,C,\lambda_f$ to the $P$-wave phase shift below $\sqrt{s} = 1.2\mbox{ GeV}$, {\it i.e.} below the energy at which inelasticity becomes sizable (since $t$ in Eq.(\ref{tqm}) describes a single channel). The result of the fit is shown in Fig.~\ref{delta} by the dashed line. Below $\sqrt{s} =  1.5\mbox{ GeV}$ the solution of the dispersion relation and the simple quark model parametrization are essentially indistinguishable. The comparison between $N$, $D$ and $t$ from dispersion relation calculation and the quark model is shown in Fig.~\ref{Re-ImN},~\ref{Re-ImD}, and \ref{Re-Imt}, respectively. 
For $s>0$ and in the $\rho$ region, $t$ computed from dispersion relations and the model should agree since both reproduce the phase shift and inelasticity. 
The $D$ function is  sensitive to the phase of the amplitude over the whole energy range and since  the phases of $t$ in both cases approach $180^0$ there is also little difference
  below the inelastic region. The only noticeable difference is in  the "potential", {\it i.e.} the numerator function $N$. In the case of the model in Eq.(\ref{tqm}) there is a double pole at $s = -s_f= -3.1\mbox{ GeV}^2 $ imposed by the simple parametrization of Eq.(\ref{fmodel}),  while 
   the actual solution of the dispersion relation shows a structure at $s \sim -1 \mbox{ GeV}^2 $ and $s \sim 2.5\mbox{ GeV}^2$.   It is clear that knowledge of the amplitude in the physical region alone
   does not allow one to reconstruct the potential. This is because the $\rho$ is dominated by the CDD pole and the "potential"  interaction, {\it i.e.}  rescattering  of the pions, has little effect on the $\rho$. This can also be seen in Fig. \ref{delta} where
   the comparison between the full solution and the   no-CDD solution is shown.

\subsection{ $\rho$ meson production} 
 \begin{figure}[h]
{\includegraphics[height=7cm]{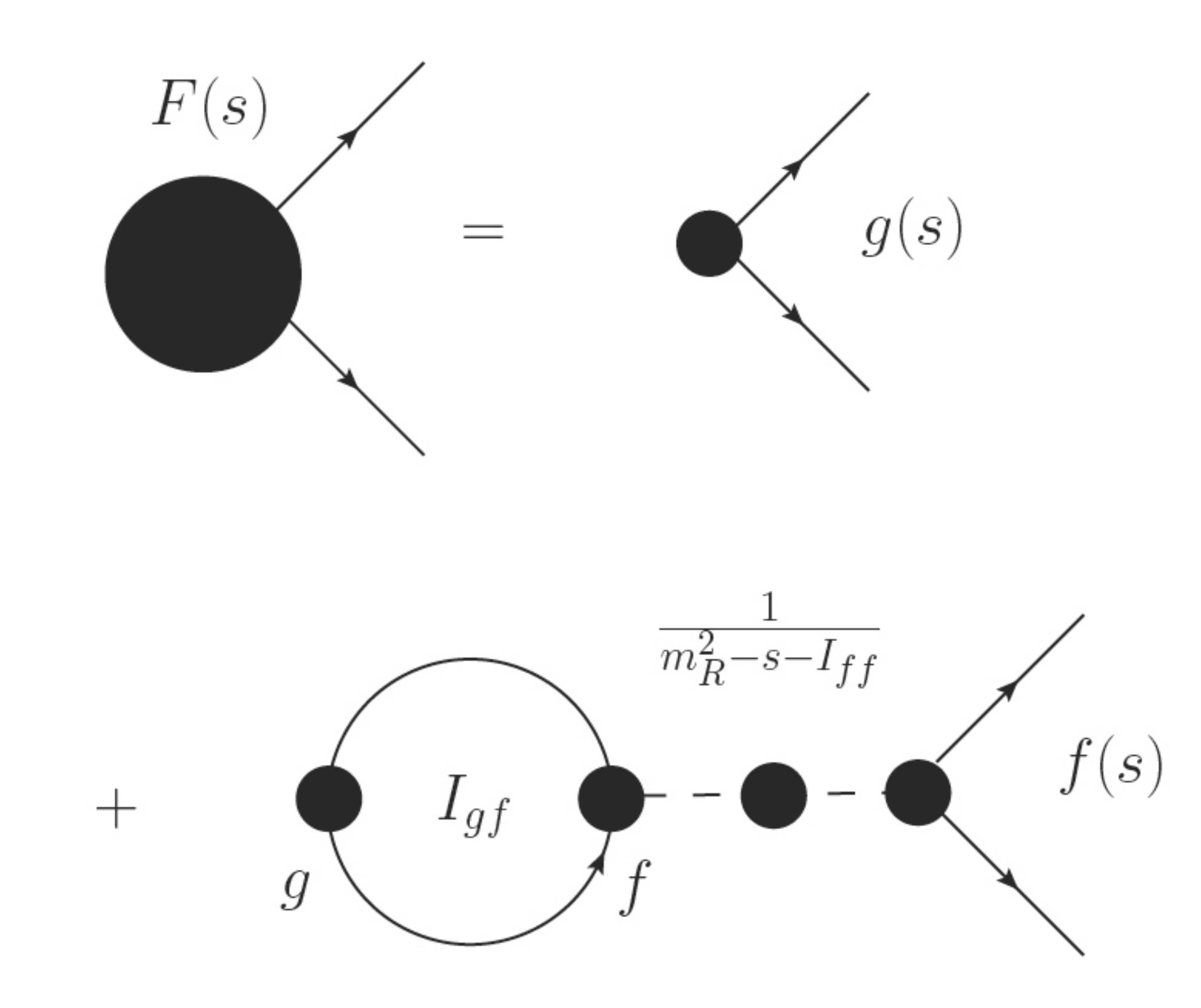}}
\caption{  Quark model representation of the $\rho$ production } 
 \label{Fqm} 
\end{figure}
\begin{figure}[h]
{\includegraphics[height=7cm]{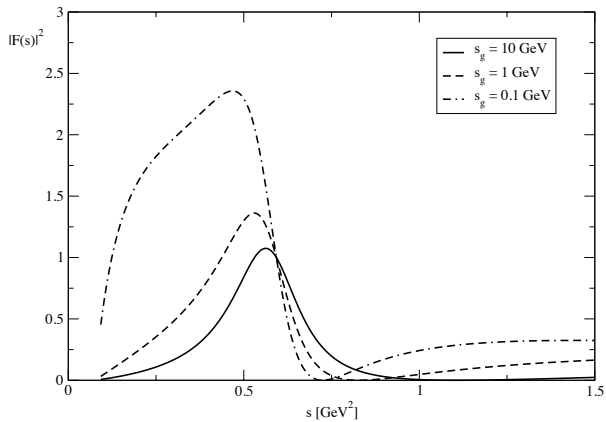}}
\caption{  Modulus squared of the production amplitude $|F(s)|$ from Eq.(\ref{Fqm1}) computed for  different values of the direct, $\pi\pi$ production range $s_g$. 
  For large value of $s_g$, {\it i.e} $s_g \gsim 1\mbox{ GeV}^2$, the interaction range is small and the shape of a quark model (compact) resonance is unaffected. 
   As the interaction range increases  the rate of low mass pion production increases distorting  resonance shape at lower energies. } 
 \label{ssprod} 
\end{figure}

We finally illustrate the role of the left hand cut on the $\rho$ shape in a specific production process. In a  quark model description  the  production amplitude $A(s) = q F(s) = (s - s_{th})^{1/2} F(s)$, ({\it c.f.} Eq.(\ref{A}))  becomes
\begin{equation} 
F(s) = g(s) + \frac{ f(s) I_{gf}(s) }{ m_B^2 - s - I_{ff}(s)},  \label{Fqm1} 
\end{equation} 
 where 
\begin{equation} 
I_{gf}(s)  = \frac{1}{\pi} \int^\infty_{s_{th}} ds_R \frac{(s_R - s_{th}) \rho(s_R) g(s_R) f(s_R)}{s_R - s}.  
\end{equation} 
and is depicted in Fig.~\ref{Fqm}. 
The first term in Eq.(\ref{Fqm1}) corresponds to direct production of the two-pions, and is given by a "potential" term, $g(s)$, which we parametrize as 
 \begin{equation} 
g(s) = \frac{\lambda_g s_g }{s + s_g} 
\end{equation} 
with $s_g > 0$ characterizing   the range of interaction or  the size  ($\sim 1/\sqrt{s_g}$) of the source.  A compact, or point-like source  corresponds to $g(s) \to const.$,
  {\it i.e.} $s_g \to \infty$  while a  spatially  extended source has $s_g \to 0$ (with $\lambda_g s_g$ fixed). The production amplitude in Eq.(\ref{Fqm1}) can be cast in the form of the general representation of Eq.(\ref{F1}). In this case $G_i(s)= 0$ since the model contains no inelastic channels, and 
   \begin{equation} 
   G(s) = C \left[ g(s) (m_B^2 - s) - \left( g(s) I_{ff}(s) - f(s) I_{gf}(s) \right)  \right]. 
   \end{equation} 
Indeed,  the right hand cut singularities of $I_{ff}$ and $I_{gf}$ cancel out and $G(s)$ has only the left hand cut  singularities of $g(s)$ and $f(s)$.  
 Similarly Eq.(\ref{Fqm1}) can be written in the form of Eq.(\ref{Wt})  with 
\begin{equation} 
Im F_L(s)  = - \pi \delta(s + s_g) \lambda_g s_g  F_g - \pi  \delta(s  + s_f)  \lambda_f s_f  F_f,
\end{equation} 
where 
\begin{eqnarray} 
F_g & = &  C \frac{ m_B^2 + s_g  - I_{ff}(-s_g) }{  D(-s_g)}  \nonumber \\
F_f &= &  C\frac{ I_{gf}(-s_f) }{  D(-s_f)} 
  \end{eqnarray}
  and so 
  \begin{equation} 
  F_L(s) =    F_g  g(s) +  F_f f(s).  \label{FLmodel} 
   \end{equation} 
In the case of a diffuse source the interaction range is small, and the principal value integral in 
Eq.(\ref{B}) is suppressed. In this case the production amplitude is suppressed at the resonance mass, since the other term is proportional to $\cos\delta$ which vanishes.  In the quark model model case,  however, we see that regardless of the range of $\pi\pi$ production,  determined  by  $g(s)$,  the production potential also contains a term 
  responsible for direct  production of the quark state, {\it i.e.} the  term proportional to $f(s)$ in 
  Eq.(\ref{FLmodel}). Thus the resonance associated with a quark state or CDD pole 
    is expected to show up as a bump in a production process. As expected, the $g$ term in 
     Eq.(\ref{FLmodel}) becomes more relevant as for  diffuse sources, 
      {\it i.e.} when $s_g << s_f$ and   $P.V. I_{gf} < P.V. I_{ff}$. This effect is illustrated in Fig.~\ref{ssprod}.

 \section{Summary and Outlook} 
 \label{sec-3} 
  
Development of analytical parameterizations of hadron production amplitudes is necessary for a successful partial wave analysis. Here we presented the analysis of the  $\pi\pi$ $P$-wave amplitude. 
The dominant feature in the elastic region is the $\rho(770)$ resonance which cannot be accounted for by "potential" interactions {\it i.e.} the  discontinuity of the amplitude across the left hand cut. Instead it seems  to originate from internal structure of the scatterers {\it i.e.} the CDD pole. The precise location of the pole cannot be determined without better knowledge of the phase shift and inelasticity in the high ($\sqrt{s} \gsim 2\mbox{ GeV}$) energy range, however for practical applications, it may be assumed that the CDD pole is at infinite energy, {\it i.e.} $\eta(\infty) \to 1$, $\delta(\infty) \to \pi$. While this is consistent with the expectation for the  asymptotic  behavior of the pion form factor~\cite{De Troconiz:2001wt,Lepage:1980fj} it contradicts the expectation that inelastic open channels dominate at high energies ($\eta(\infty) \to 0$) and requires further studies. The CDD pole in a partial wave amplitude is equivalent to presence of a "bare" state {\it e.g.} quark bound state, which couples to the $\pi\pi$ continuum. It is thus not surprising that a generic quark model description of $P$-wave $\pi\pi$ scattering with a single bound state faithfully describes the $P$-wave amplitude up to $\sqrt{s} \sim 1.6 \mbox{ GeV}$ where  inelasticity  becomes significant indicating a large $K{\bar K}$ 
 component of the $\rho'(1600)$. While the $\rho$ resonance shape can  be distorted if the production amplitude varies strongly near threshold, to our best knowledge 
 such features has not been observed in  the experimental data. Even though  the left hand cut behavior of the $P$-wave amplitude in scattering and production does not seem to drive the $\rho$ resonance, this  does not need to be the case for other resonances ({\it e.g.} the sigma meson)
  leading to the conclusion that  the left hand cut properties of amplitudes require attention  in partial wave analyses.

\section{ACKNOWLEDGMENTS  } 
We would like to thank Christoph Hanhart, Mark Paris and Jose Pelaez  for numerous discussions. 
This work was supported in part by the US Department of Energy grant under 
contract DE-FG0287ER40365. 
  We also acknowledge partial support 
  Instituto Nazionale di Fisica Nucleare,   Sezione di Genova, 
      that allowed his collaboration to continue.


\begin{thebibliography}{99}

 
  
    
  
 
 \bibitem{PDG} C. Amsler {\it et al.}, Phys. Lett. B{\bf667}, 1 (2008). 



\bibitem{Oller:1997ng}
J.~A.~Oller, E.~Oset and J.~R.~Pelaez,
Phys.\ Rev.\ Lett.\  {\bf 80} (1998) 3452.



\bibitem{Oller:1998zr}
J.~A.~Oller and E.~Oset,
Phys.\ Rev.\ D {\bf 60} (1999) 074023.

  
  \bibitem{Pelaez:2003eh}
  J.~R.~Pelaez and F.~J.~Yndurain,
  Phys.\ Rev.\  D {\bf 68}, 074005 (2003).
  


\bibitem{Caprini:2005zr}
 I.~Caprini, G.~Colangelo and H.~Leutwyler,
 Phys.\ Rev.\ Lett.\  {\bf 96}, 132001 (2006)

   
\bibitem{Kaminski:2006qe}
  R.~Kaminski, J.~R.~Pelaez and F.~J.~Yndurain,
  Phys.\ Rev.\  D {\bf 77}, 054015 (2008).

 
\bibitem{Caprini:2008fc}
  I.~Caprini,
  Phys.\ Rev.\  D {\bf 77}, 114019 (2008).
  
  
  
  
\bibitem{Hanhart:2008mx}
 C.~Hanhart, J.~R.~Pelaez and G.~Rios,
 Phys.\ Rev.\ Lett.\  {\bf 100}, 152001 (2008).



\bibitem{Pelaez:2003dy}
 J.~R.~Pelaez,
 Phys.\ Rev.\ Lett.\  {\bf 92}, 102001 (2004).
\bibitem{Pelaez:2006nj}
 J.~R.~Pelaez and G.~Rios,
 Phys.\ Rev.\ Lett.\  {\bf 97}, 242002 (2006).



\bibitem{GarciaRecio:2003ks}
  C.~Garcia-Recio, M.~F.~M.~Lutz and J.~Nieves,
  Phys.\ Lett.\  B {\bf 582}, 49 (2004). 

\bibitem{Guo}
F.~K.~Guo, C.~Hanhart, F.~J.~Llanes-Estrada and U.~G.~Meissner,
 Phys.\ Lett.\  B {\bf 678} (2009) 90.



\bibitem{Aitchison:1970tp}
  I.~J.~R.~Aitchison and P.~R.~Graves-Morris,
  Nucl.\ Phys.\  B {\bf 14}, 683 (1969).


\bibitem{Aitchison:1969tq}
  I.~J.~R.~Aitchison and C.~Kacser,
  Phys.\ Rev.\  {\bf 173}, 1700 (1968).
  
  
  \bibitem{Aitchison:1977sm}
  I.~J.~R.~Aitchison and M.~G.~Bowler,
  J.\ Phys.\ G {\bf 3}, 1503 (1977).


\bibitem{Bowler:1975my}
  M.~G.~Bowler, M.~A.~V.~Game, I.~J.~R.~Aitchison and J.~B.~Dainton,
  Nucl.\ Phys.\  B {\bf 97}, 227 (1975).

  
\bibitem{Pennington:2006dg}
  M.~R.~Pennington,
  Phys.\ Rev.\ Lett.\  {\bf 97}, 011601 (2006).
  
  
    
\bibitem{Battaglieri:2008ps}
 M.~Battaglieri, R.~De Vita, A.~P.~Szczepaniak and f.~t.~C.~Collaboration,
  Phys.\ Rev.\ Lett.\  {\bf 102}, 102001 (2009).
  
\bibitem{Battaglieri:2009fq}
  M.~Battaglieri, R.~De Vita, A.~P.~Szczepaniak and f.~t.~C.~Collaboration,
  arXiv:0907.1021 [hep-ex].

\bibitem{CLEO} P.Guo, A.Szczepaniak, M.Shepherd, R.Mitchell, {\it in preparation.} 

\bibitem{Truong:1988zp-2}
T.~N.~Truong,
Phys.\ Rev.\ Lett.\  {\bf 67}, (1991) 2260. 


\bibitem{Dobado:1996ps}
A.~Dobado and J.~R.~Pelaez,
Phys.\ Rev.\ D {\bf 47} (1993) 4883.

\bibitem{Dobado:1996ps-2}
A.~Dobado and J.~R.~Pelaez, 
Phys.\ Rev.\ D {\bf 56} (1997) 3057.

   
\bibitem{CM}   
  G.F. Chew and S. Mandelstam, Phys. Rev. {\bf 119}, 467  (1960).


\bibitem{Bjorken:1960zz}
  J.~D.~Bjorken,
  Phys.\ Rev.\ Lett.\  {\bf 4}, 473 (1960).
  
 \bibitem{FW} 
   G. Frye and R.L. Warnock, Phys. Rev. {\bf 130}, 478 (1963). 


\bibitem{CDD}  
   L. Castillejo, R.H. Dalitz and F.J. Dyson, Phys. Rev. {\bf 101}, 453  (1956).
   
   
        \bibitem{Lev} Phase shifts are defined modulo $\pi$ and we use the conventional normalization 
      $\delta(s_{th}) =0 $. Furthermore we have assumed there are no $\pi\pi$ bound states, which 
       with this phase shift normalization would appear as explicit zeros in the {\it rhs} of 
        Eq.(\ref{OO}).

     
\bibitem{M} 
N.I. Muskhelishvili, Tr. Tbilis. Math Instrum. {\bf 10}, 1 (1958); in {\it Singular Integral Equations}, J.Radox, ed. (Noordhoff, Groningen, 1985). 

\bibitem{O}
R. Omn\'es, Nuovo Cim. {\bf 8}, 316 (1958). 

\bibitem{Pham:1976yi}
  T.~N.~Pham and T.~N.~Truong,
  Phys.\ Rev.\  D {\bf 16}, 896 (1977).
  
\bibitem{Yndurain:2002ud}
  F.~J.~Yndurain,
  arXiv:hep-ph/0212282.
  
    \bibitem{Tryon:1974tq}
  E.~P.~Tryon,
  Phys.\ Rev.\  D {\bf 12}, 759 (1975).


  
    \bibitem{Hyams:1973}
B. Hyams, C. Jones and P. Weilhammer,
  Nucl.\ Phys.\  B {\bf 64}, 134(1973).
  
   \bibitem{Protopopescu:1973}
S.D. Protopopescu {\it et al.} ,
  Phys.\ Rev.\  D {\bf 7}, 1279(1973).

       \bibitem{Estabrooks:1974}
P. Estabrooks and A. D. Martin,
  Nucl.\ Phys.\  B {\bf 79}, 301(1974).
  
    \bibitem{De Troconiz:2001wt}
  J.~F.~De Troconiz and F.~J.~Yndurain,
  Phys.\ Rev.\  D {\bf 65}, 093001 (2002)

\bibitem{Lepage:1980fj}
  G.~P.~Lepage and S.~J.~Brodsky,
  Phys.\ Rev.\  D {\bf 22}, 2157 (1980).

  
  

    
     
\end{thebibliography}
 \end{document}